\documentclass[aps, prd, preprintnumbers,nofootnoteinbib,notitlepage]{revtex4-1}

\usepackage{colortbl}
\usepackage[T1]{fontenc}
\usepackage{microtype}
\usepackage[textsize=scriptsize,backgroundcolor=red!70,linecolor=red]{todonotes}
\usepackage{booktabs}
\usepackage{dcolumn}
\usepackage{amsmath}
\usepackage{amsfonts}
\usepackage{amssymb}
\usepackage{graphicx}
\usepackage{hyperref}
\usepackage[all]{hypcap}
\usepackage{paralist}
\usepackage{multirow}

\usepackage{graphicx}
\usepackage{dcolumn}
\usepackage{bm}
\usepackage{epsf}
\usepackage{dcolumn}
\def\be{\begin{equation}}
\def\ee{\end{equation}}
\def\bea{\begin{eqnarray}}
\def\eea{\end{eqnarray}}
\def\gsim{\ \rlap{\raise 2pt\hbox{$>$}}{\lower 2pt \hbox{$\sim$}}\ }
\def\lsim{\ \rlap{\raise 2pt\hbox{$<$}}{\lower 2pt \hbox{$\sim$}}\ }
\def\dslash{\kern-4pt \not{\hbox{\kern-2pt $\partial$}}}
\def\pslash{\not{\hbox{\kern-2pt p}}}

\def\l{{\rm L}}

\definecolor{gray}{rgb}{0.90,1,1}
\definecolor{LightCyan}{rgb}{0.88,1,1}

\def \be{\beta}

\def\beq{\begin{equation}}
\def\eeq{\end{equation}}
\def\bea{\begin{eqnarray}}
\def\eea{\end{eqnarray}}
\def\ber{\begin{eqnarray*}}
\def\eer{\end{eqnarray*}}
\def\bwt{\begin{widetext}}
\def\ewt{\end{widetext}}

\def\roughly#1{\mathrel{\raise.3ex\hbox
{$#1$\kern-.75em\lower1ex\hbox{$\sim$}}}}
\def\lsim{\roughly<}
\def\gsim{\roughly>}

\def\order{\lower 1.8ex \hbox{\LARGE\~{}}}

\usepackage{bm}

\def \({\left(}
\def \){\right)}
\def \[{\left[}
\def \]{\right]}
\def \l|{\left|}
\def \r|{\right|}

\def \be{\beta}

\begin{document}
\DeclareGraphicsExtensions{.eps,.ps}


\title{ MSSM vs. NMSSM in $\Delta F=2$ transitions}



\author{Jacky Kumar}
\affiliation{Department of High Energy Physics, Tata Institute of Fundamental Research, Mumbai 400005, India}

 

\begin{abstract}
 We study deviations between MSSM and NMSSM in
the predictions of $\Delta F=2$ processes. We found that 
there can be two sources which can cause such deviations,
 \emph{i.e}, due to certain neutralino-gluino cross box diagrams and due to
well known double penguin diagrams. Both are effective at large $\tan \beta$.
In addition to this, taking into account 8 TeV direct search constraints 
from the heavy Higgs searches, we study the maximum allowed MFV 
like new physics (NP) effects on $\Delta M_s$ in the two models. 
In NMSSM such NP effects can be as large as $25 \%$, on the other
hand in MSSM such large contributions are severely constrained. 
\end{abstract}
\maketitle



\section{Setup}
The next-to-minimal supersymmetric standard model (NMSSM) 
\cite{Fayet:1974pd,Ellwanger:2009dp} is known for 
providing a solution to the $\mu$-problem\cite{Kim:1983dt} and accommodating a 
125 GeV SM-like Higgs boson with relatively
lesser fine tuning as compare to MSSM\cite{Hall:2011aa}. 
In addition to two Higgs doublets in
MSSM, it contains an extra gauge singlet chiral superfield($S$). 
The superpotential of ${Z}_3 $-invariant NMSSM is given by,
\begin{equation}
\mathcal{ W} = \lambda~ H_u.  H_d ~  S ~+~ \frac{\kappa}{3} ~  S^3 ~+~ \mathcal{W}_{Yukawa}.
\label{superpot}
\end{equation}
Where $\mathcal{W}_{Yukawa}$ contains the Yukawa terms and $\lambda$, $\kappa$ are 
the dimensionless couplings. 
NMSSM has an extended Higgs and neutralino sectors as compare to MSSM. 
For example the physical neutralino 
states are mixtures of gaugino, Higgsino and '\emph{singlino}'- the fermion component
of the superfield $S$, weak eigen states. 
Although the stop sector, which has an additional source of quark flavor 
violation in supersymmetry(SUSY), remains unchanged. 
But there can still be additional effects of NMSSM on the quark flavor 
transitions due its modified neutralino and Higgs  sectors.
In this regard we study the $\Delta F=2$ transitions in MSSM and 
NMSSM and address the following two questions,
\begin{itemize}
\item Can NMSSM, without any assumption on squark flavor structure, 
give different theoretical predictions for the $\Delta F=2$
observables as compare to MSSM?. If so, what is the mechanism behind this.
\item What are the largest allowed minimal flavor violating(MFV) 
effects\cite{DAmbrosio:2002vsn} in MSSM and NMSSM 
in $\Delta F=2$ observables at low $\tan \beta$?
\end{itemize}
First, we isolate all possible genuine NMSSM 
contributions(non-MSSM) to $\Delta F=2$ transitions and figure out 
the conditions in which these two models give different predictions. 
The amplitude for $B -\overline B$ mixing 
is defined as $M_{12}^q = \left < B_q|H_{eff}| \overline{B_q} \right > $,
where $q=d,s$ stand for $B_d,B_s$ mixing, respectively.
The effective Hamiltonian, $H_{eff}$, can be consistently expressed 
in the basis of eight dimension-six operators $Q_i$ as,
\begin{equation}
H_{eff} = \sum_{i} C_i Q_i +  \emph{h.c},
\label{eq:heff}
\end{equation}
with $C_i$ being their respective Wilson Coefficients (WC). We follow the operator basis defined in \cite{Buras:2002vd}, which reads explicitly,
\begin{equation}
Q^{VLL} =(\bar b_L \gamma_{\mu} q_L)(\bar b_L \gamma^{\mu}  q_L) \ , \quad  
Q^{VLR}=(\bar b_L \gamma_{\mu}  q_L)(\bar b_R \gamma^{\mu}   q_R) \ , \quad  
Q^{SLR}=(\bar b_R q_L)(\bar b_L q_R)  \ , \quad  \nonumber
\end{equation}
\begin{equation}
Q_1^{SLL}=(\bar b_R q_L)(\bar b_R q_L) \ , \quad  
Q_2^{SLL}=(\bar b_R \sigma_{\mu \nu} q_L)(\bar b_R \sigma^{\mu \nu} q_L) \ . \quad  
\label{bbops}
\end{equation}
\begin{figure}[h]
\centering
  \includegraphics[width=12cm,trim={3cm 18cm 0 3.8cm},clip]{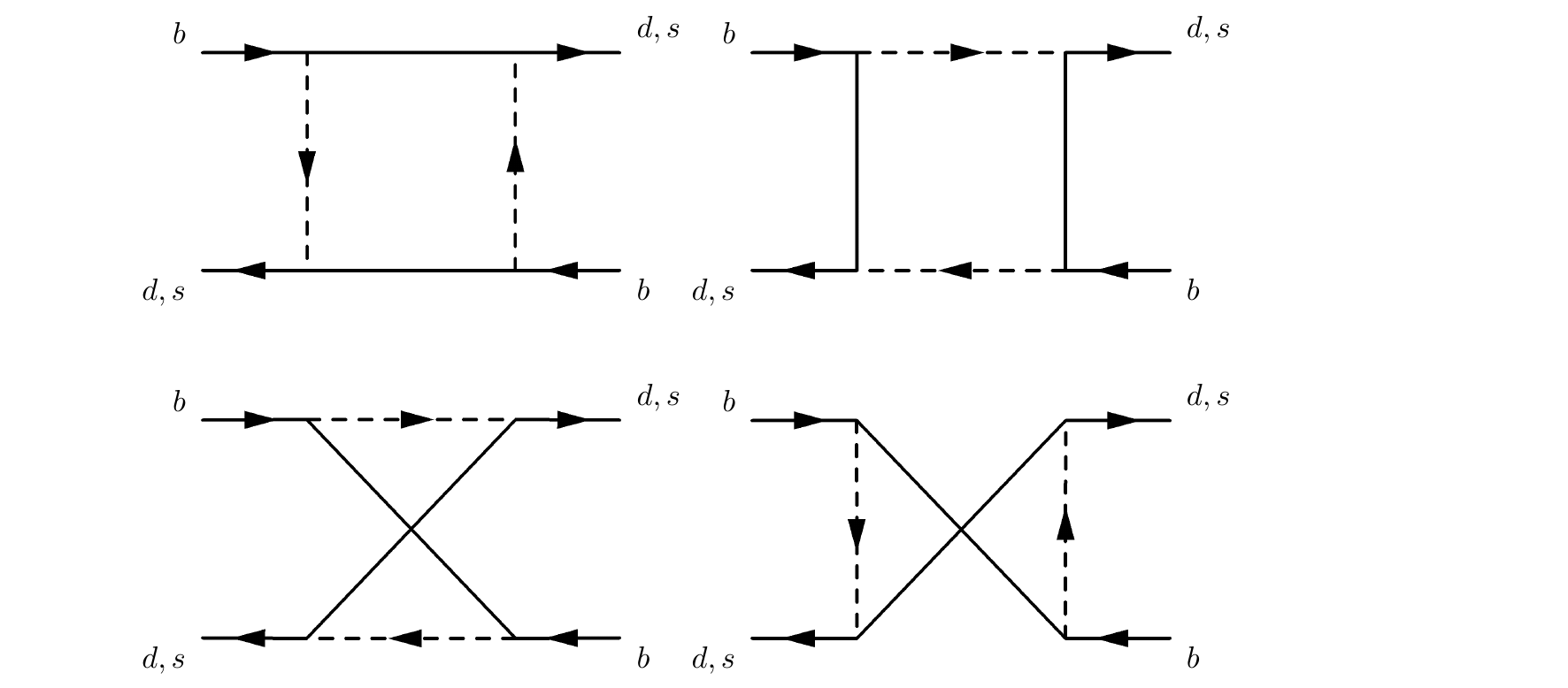}
  \caption{\small One-loop box diagrams contributing to $\Delta F=2$ observables.}
\label{fig:mssm}
\end{figure}\normalsize
In the above expressions, the diagonal quark-color indices are suppressed (assumed to be contracted separately within each bracket), and $\sigma_{\mu \nu} =\frac{1}{2}[\gamma_{\mu},\gamma_{\nu}]$. The remaining three operators,
$Q^{VRR}, Q_1^{SRR}, Q_2^{SRR}$  are obtained from
$Q^{VLL}, Q_1^{SLL}, Q_2^{SLL}$ by
interchanging $L$ with $R$.
In SM only $Q^{VLL}$ gets non-zero contribution from 
one-loop \emph{box diagrams} with quarks
and $W$-bosons circulating in the loops. But in MSSM 
there are various, additional, box contributions  
mediated by: 1) charged Higgs, up-quarks; 2) chargino, 
up-squarks; 3) gluinos, down-squarks; 4) neutralinos, 
down-squarks; 5) gluino, neutralino, down 
squarks \cite{Altmannshofer:2007cs}. Their diagrammatic 
topologies are shown in Fig.\ref{fig:mssm}. 
Certain two-loop diagrams (\emph{i.e., double-penguins}) 
which depend on positive powers of $\tan\beta$ become also 
relevant for large values of this parameter and can easily dominate over any other contribution \cite{Buras:2002vd}.

\subsection{NMSSM contributions to $\Delta M_{s}$}
There can be two kind of genuine NMSSM effects, 1) due to presence 
of an extra neutralino state 2) due to additional Higgs bosons.
The extra neutralino state trivially affects the neutralino-
quark-squark vertex(V), which appears in the box diagrams shown in Fig.\ref{fig:mssm}.
But this can leave strong imprints on the observables. However we 
find that the largest new effects arise only due to the crossed box diagrams 
mediated by neutralino and gluino.
A typical WC originating from neutralino-gluino contributions 
in the mass basis has a from
(see Appendix A of \cite{Kumar:2016vhm} for all WCs),
\begin{eqnarray}
C_{1}^{SRR} & =& \frac{g_3^2}{16 \pi^2}\frac{7}{6} 
~V^R_{2ka} ~V^{L*}_{3la} ~Z^*_{3k}  Z_{5l} 
~ m_{\tilde g} ~m_a ~D_0(m_{\tilde g}^2,m_{a}^2,m_k^2,m_l^2) \nonumber \\
 ~ & +& \frac{g_3^2}{16 \pi^2} ~  \frac{1}{6} ~ 
\left (V^{L*}_{3ka} ~V^{L*}_{3la} ~ Z_{5k}~ Z_{5l} ~ +~ V^R_{2ka} ~V^R_{2la} 
~Z^*_{3k}~Z^*_{3l}\right)   ~m_{\tilde g} ~m_a ~
D_0(m_{\tilde g}^2,m_{a}^2,m_k^2,m_l^2).     
\label{eq:wc}
\end{eqnarray}
Since singlino mixes only with the Higgsinos via first term in the 
superpotential Eq(\ref{superpot}), so keeping only Yukawa related terms in the 
vertex (V), we identify two types basic structures appearing in all the WCs,
\begin{eqnarray}
& m_{\tilde g} ~(Z_N)_{3a} ~m_a D_0(m_{\tilde g}^2 , m_a^2, x) ~(Z_N)_{3a}\label{eq:puremass}\\
&(Z_N)_{3a} ~D_2(m_{\tilde g}^2, m_a^2,x) ~(Z_N)_{3a}^*\label{eq:mixedmass}
\end{eqnarray}
where the other factors like $g_3^2Y_b^2$, $Z_D$ are suppressed and the two 
down-squark mass arguments of the loop-functions are suppressed into the argument $x$. 
Up to complex conjugation in the above expressions, one can easily verify that there is 
no other structure \cite{Dedes:2015twa}.

\begin{figure}[t]
\includegraphics[height=6.0cm,width=8.6cm]{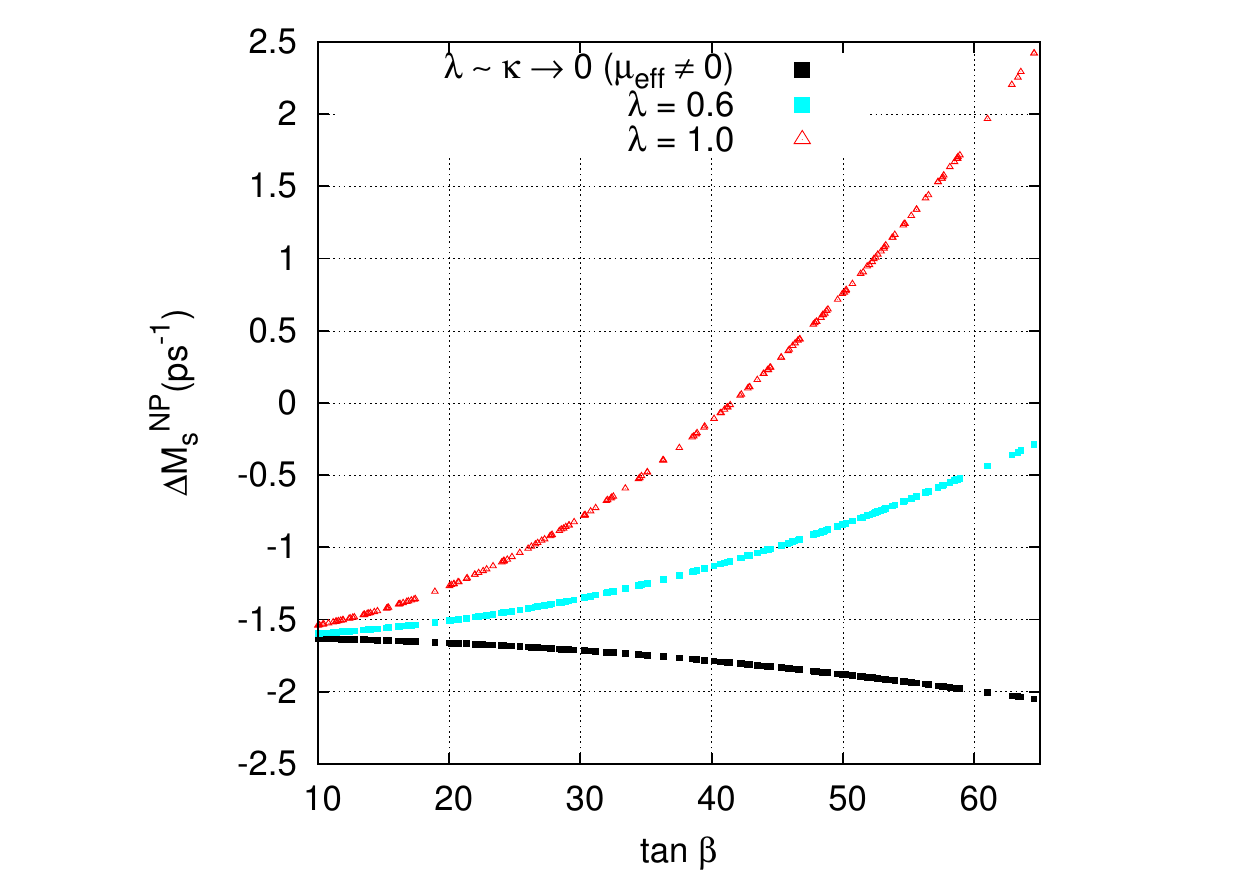}\hspace{-2.7cm}
\includegraphics[height=5.8cm,width=9.1cm,trim={0.cm 0.37cm 0.cm 0.3cm},clip]{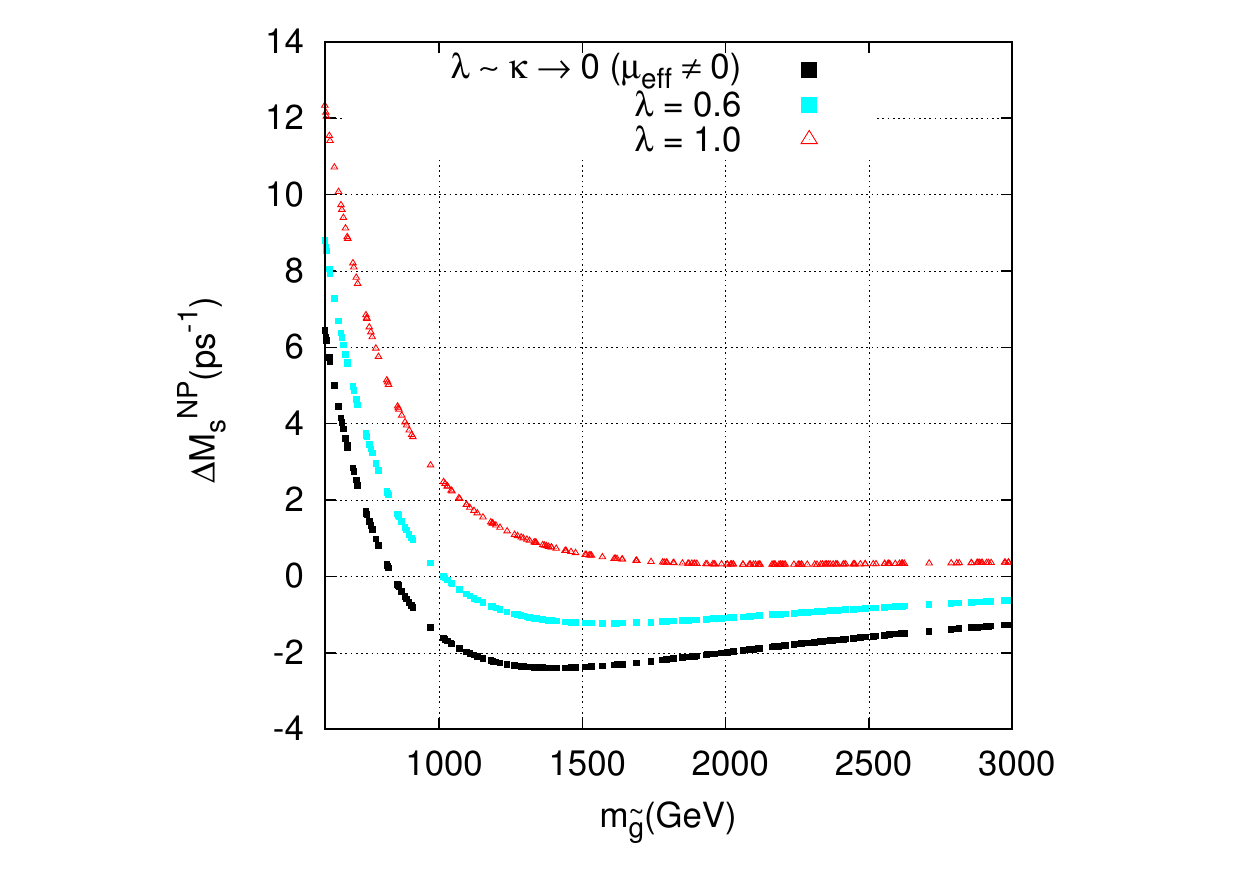}
\caption{\small  Genuine-NMSSM effects in $\Delta M_s$, understood as deviations with respect 
to the MSSM predictions under $\tan\beta$ (left) and gluino mass (right), scaling. 
Input parameters primarily controlling the effect 
read $(m_D^2)_{ii}=650~GeV, M_S=3 ~TeV, \delta_{RR}^{23}=0.6 $, 
while $m_{\tilde{g}}=1.1~TeV$ and $\tan\beta =60$ were used 
for left and right plot, respectively. Cyan line ($\kappa=0.4$) 
corresponds to perturbative NMSSM up to GUT-scale. 
Red line ($\kappa=1$) requires UV-completion before 
GUT-scale, as in $\lambda$-SUSY models. The black line 
is the MSSM-limit of the NMSSM model. For other parameters 
see text. Calculations are performed in mass basis taking into account all contributions.}\label{fig:dms}
\end{figure}

Next, to isolate the genuine NMSSM contributions, 
we use (Flavor expansion theorem)FET\cite{Dedes:2015twa},
and translate the mass eigen state expressions of Eq(\ref{eq:wc}) 
into the Mass Insertion Approximation(MIA) expansion, i.e. 
\small\begin{eqnarray}
 m_{\tilde g} ~ \Big[\mathbf{M_N} D_0(m_{\tilde g}^2,\mathbf{M_N^2},x)\Big]_{33}&=&
 m_{\tilde g} ~ (M_N)_{35}~(M_N^2)_{53}~ E_0\Big(m_{\tilde g}^2,(M_N^2)_{55} ,(M_N^2)_{33},x \Big)+\dots\label{eq:pureflav}\\
 \Big[D_2(m_{\tilde g}^2, \mathbf{M_N^2},x)\Big]_{33}&= &~D_2\Big(m_{\tilde g}^2, (M_N^2)_{33},x\Big)+\dots \label{eq:mixedflav}
\end{eqnarray}\normalsize
where the dots represent the terms higher order in the 
neutralino mass insertions(higher order in FET).
The explicit form of all loop functions 
$D_0,~ D_2$ and $E_0$ can be found in \cite{Kumar:2016vhm}.
The leading \emph{genuine-NMSSM effects} come from $E_0$-terms, 
having a strong dependence on $\lambda,\kappa$-parameters 
through neutralino mass matrix elements $(M_N)_{35}$ and $(M_N^2)_{53}$ 
which are, in addition, related to $v_u$. Although  suppressed by a 
neutralino mass insertion these can be important 
when Higgsino-singlino, i.e. $\tilde H_d^0 - \tilde S$ 
mixing is sufficiently large. 
The $D_2$-terms are less sensitive to the NMSSM parameters $\lambda,\kappa$ 
since these appear only through the $(M_N^2)_{33}$ argument of 
the respective loop function. In this sense, $D_2$-terms  
mediate \emph{mixed effects} which is understood by the 
fact that they are non-zero in the MSSM limit, $\lambda\sim\kappa\to 0$.  
Typically, the $E_0$-terms are safe from $D_2$-term screening, since 
they are primarily associated with different types of squark mass 
insertions. Nevertheless, due to neutralino mass insertion 
suppression, the $E_0$-term can become comparable to other 
neutralino-gluino MSSM contributions. These are sub-leading 
in the couplings (\emph{e.g.,}$\propto Y_b Y_s,g_2^2$,etc.) 
but not suppressed by neutralino insertions. In 
Fig.\ref{fig:dms}, we present the relative magnitude of 
genuine-NMSSM and MSSM contributions. Left panel 
shows $\Delta M_s$ as a function of $\tan\beta$, while right 
plot shows the variation of same with gluino mass. The qualitative 
analysis above dictates following general properties of NMSSM enhanced region.
\normalsize
Large values of $\tan\beta$ and $\lambda\sim\kappa$ are required. 
The former condition enhances the down-type Yukawa couplings  
which are present in $\tilde H_d^0$ interactions. The 
latter condition is required for large Higgsino-singlino 
mixing which controls the size of genuine-NMSSM contributions. 
Typical values for significant effects 
are $50\lesssim \tan\beta~ (\lesssim 65)$ 
and $0.5\lesssim \kappa\sim\lambda~(\lesssim 1) $. 
Large values of $M_A$ are preferable, which suppress both 
charged Higgs contributions and double penguins effects. 
This is also motivated by the Higgs potential in the 
large $\tan\beta,\lambda$ regime of NMSSM, as discussed 
in appendix C of \cite{Kumar:2016vhm}. There, we display 
the method of obtaining phenomenologically viable 
CP-even and CP-odd scalar masses by fitting 
the soft $A_\lambda, A_\kappa$ parameters, 
while keeping $\lambda,\kappa$ as free parameters. 
The typical range for $M_A$  obtained this 
way is $4 ~TeV\lesssim M_A\lesssim 12 ~TeV$, 
depending on $\mu_{eff},\tan\beta$ inputs.

Another source of genuine NMSSM effects is double penguin diagrams. 
The situation in this case is more involved and although 
the effect of an extra neutralino circulating in loops 
is in practice irrelevant, the extra CP-even and CP-odd 
singlet states induce various modifications in relevant 
couplings and spectra. NMSSM effects in double penguin 
diagrams are effective mostly when mass of lighter 
pseudoscalar is around 5 GeV (which is the B-meson mass). 
This is basically a resonance effect. This effect can 
cause even an order of magnitude enhanced NP 
contributions at the resonance and is effective 
only at large $\tan \beta$\cite{Kumar:2016vhm}.
\subsection{Upper bounds on new physics in $\Delta F=2$ for 
MFV models at $\tan \beta$ in MSSM and NMSSM}
\begin{figure}[t]
\centering
  \includegraphics[width=0.6\textwidth]{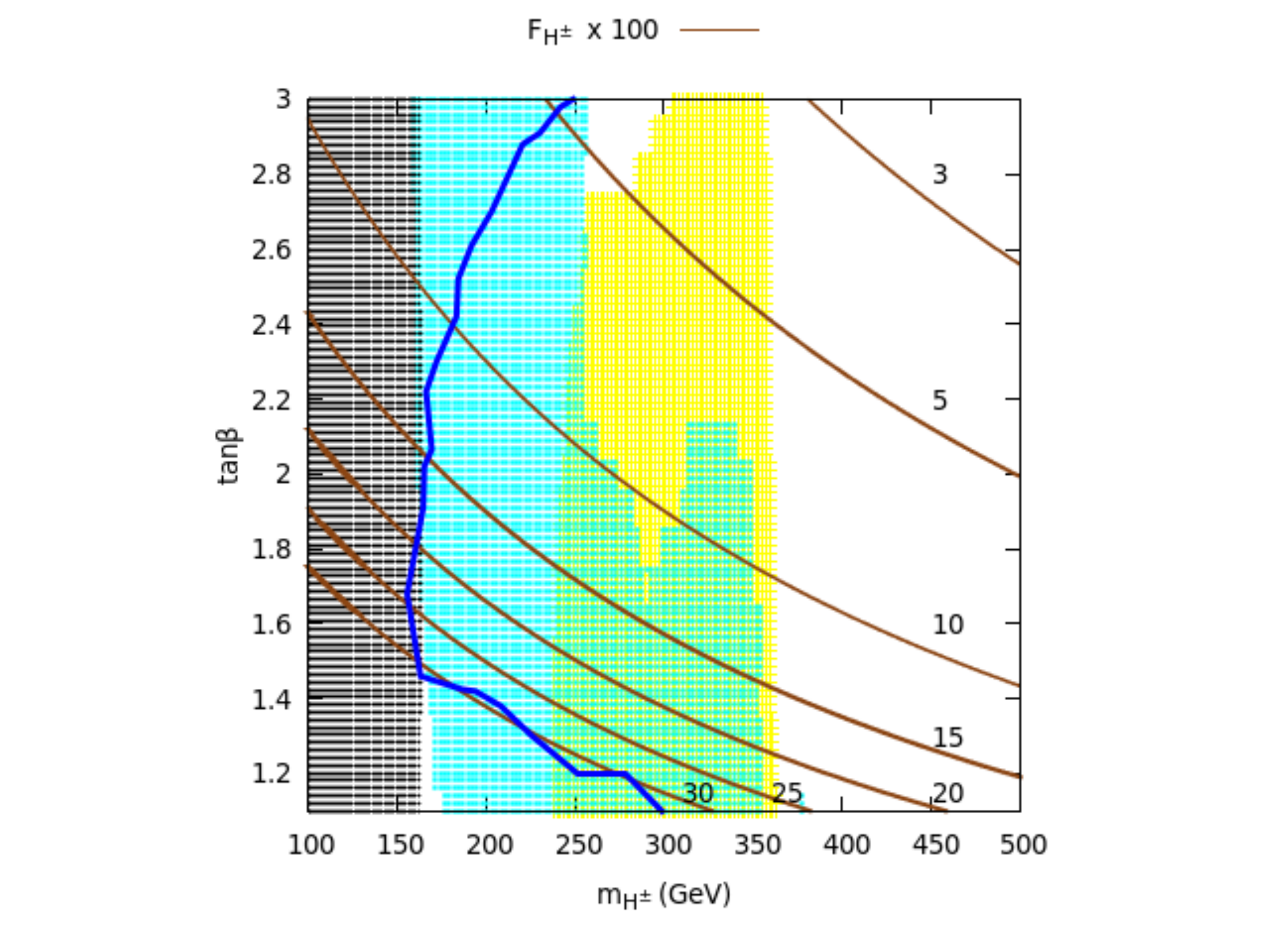}
\caption{\small Brown contours shows percentage modification
$F_{H^{\pm}}$ to $\Delta F=2$ observables involving charged Higgs.
Gray $(H^+ \rightarrow \tau^+ \nu)$, cyan $(H \rightarrow ZZ)$ and yellow $(A \rightarrow hZ)$ regions are hMSSM exclusions at $95 \% CL$. NMSSM exclusion is on the left-side of the blue contour.}
\label{ma-tb-mssm}
\end{figure}\normalsize
From our previous analysis we conclude that at low $\tan \beta$ NMSSM 
gives same predictions for $\Delta F=2$ transitions irrespective 
of MFV assumption. It is to be noted that common SUSY-parameter 
of both models has to lie in physical parameter space. 
After Higgs discovery MSSM at low $\tan \beta$ 
survives only through hMSSM scenario\cite{Djouadi:2013uqa}, which requires a 
very high SUSY scale to reach to 125 GeV Higgs mass at low $\tan \beta$.
It is known that with the MFV assumption once we take into 
account the direct search bounds on sparticles, the 
dominant contribution comes from charged Higgs diagrams\cite{Barbieri:2014tja}.
This simplifies the picture a lot, because charged Higgs 
contributions are mainly controlled by two 
parameters $M_A$ and $\tan \beta$. On the other hand 
direct searches at the LHC has set stringent bounds on 
the these two parameter. Particularly the searches 
sensitive to low $\tan \beta$ include 
$H \rightarrow ZZ$ \cite{ATLAS:2016npe},~ 
$H^+ \rightarrow \tau \nu$\cite{Aaboud:2016dig}, 
$A \rightarrow hZ$\cite{TheATLAScollaboration:2016loc}. 
We employ 8TeV data in these channels to set limits on the 
charged Higgs mass as  a function of $\tan \beta$ in both the models.
Using these limits we set upper bound on NP in $\Delta F=2$ 
observables. This is shown in Fig. \ref{ma-tb-mssm}. The 
brown contours represent the percentage deviations $F_{H^{\pm}}$ 
in $\Delta F=2$ 
due to charged Higgs contribution. Clearly $O(25) \%$ 
contribution is MSSM is severely constrained, on the other 
hand in NMSSM the situation is more relaxed. This means 
for NMSSM, at present constraints on the $M_A- \tan \beta $ plane
 coming from flavor physics sector 
are comparable or stronger than direct searches.
\section{Summary}
We study NMSSM contributions to $\Delta F=2$
transitions and find that such effects can come 
either from certain neutralino-gluino crossed box diagram,
due to the extended neutralino sector of NMSSM, and from 
double-penguin diagrams due to the extra scalar states, 
both are effective in the large $\tan \beta$ regime. We also study the 
low $\tan \beta$ regime, where a distinction between these two models in $\Delta F=2$ 
processes can come indirectly, due to different constraints on the 
allowed parameter space of the two models. To this end, we incorporate 
the recent limits from $H \rightarrow ZZ,~ A\rightarrow hZ$ 
and $H^{\pm} \rightarrow \tau \nu$ along with Higgs observables and 
set upper bounds on the new physics contributions of the two models 
under the MFV assumption. We find that an $O(25 \% )$ contribution 
in $\Delta F$ = 2, originating from charged-Higgs diagrams is 
still possible in both models, however such a large effect is 
severely constrained in the case of MSSM due to stronger bounds on the charged Higgs mass.

\bigskip
\noindent
{\bf Acknowledgements}: I would like to thank the organizers of CKM 2016 
workshop for giving an opportunity to present this work. Also, thanks to Michael 
paraskevas for a pleasant collaboration in this work.

\end{document}